\documentclass[prl,aps,showpacs]{revtex4}

\usepackage{epsf}
\usepackage{bm}
\usepackage{multirow}
\usepackage[dvips]{graphicx}

\begin{document}

\title{From graphene to {\it graphane} : A density functional investigation of metal insulator transition}
\author{Prachi Chandrachud$^1$, Bhalchandra S. Pujari$^1$, and D. G. Kanhere$^{1,2}$}
\affiliation {$^1$Department of Physics, 
         University of Pune, 
         Ganeshkhind, 
         Pune--411 007, 
         India.\\
$^2$Centre for Modelling and Simulations, University of Pune,  Ganeshkhind,
 Pune--411 007, India. 
}
\begin{abstract}
  
  While graphene is a semi-metal, recently synthesized hydrogenated graphene called {\it
  graphane}, turns out to be an insulator.  We have probed the metal insulator Transition
  in graphene-{\it graphane} system within the framework of density functional theory.  By
  analysing the evolutionary trends in the electronic structure for fifteen different
  hydrogen concentrations on graphene, we unravel some novel features of this transition.
  As hydrogen coverage increases the semi-metal turns first into a metal, then transforms
  into an insulator.  The metallic phase is spatially inhomogeneous in the sense, it
  contains the islands of insulating regions formed by hydrogenated carbon atoms and the
  metallic channels formed by contagious naked carbon atoms. 

\end{abstract}

\pacs{61.48.Gh, 81.05.ue}
\maketitle

Carbon turns out to be one of the most versatile elements forming a wide variety of
structures such as three dimensional sp$^{3}$ bonded solids like diamond, sp$^{2}$
hybridized two dimensional systems like graphene and novel nano structures like fullerenes
and nanotubes.  The electronic structure and the physical properties of these carbon based
materials are turning out to be exotic.\cite{smalley}  Although the existence and the
properties of three dimensional allotrope, graphite, containing weakly coupled stacks of
graphene layers were well known,\cite{wallace} the experimental realization of a monolayer
graphene, brought forth a completely different set of novel properties.  \cite{gaim,
castro-phys} The triangular bipartite lattice of graphene leads to the electronic
structure with linear dispersion near the Fermi points.  The low energy behavior of
two dimensional electrons in graphene has been a subject of intense experimental and
theoretical activity exploring electronic, magnetic, structural and other properties.  For
a recent review the reader is referred to Castro Neto {\it et al.}~\cite{castro}

Although graphene is considered as a prime candidate for many applications, the absence of
a band gap is a worrisome feature for the applications to the solid state devices. In the
past, several routes have been proposed to open a band gap.  \cite{route1,route2,route3}
The most interesting one is the recent discovery of completely hydrogenated graphene sheet
named as {\it graphane}.\footnote{ The reported band gap for {\it graphane} is 3.5 eV and
5.4 eV using DFT and GW-based calculations respectively.\cite{sofo,GW} There are
no experimental estimates.} {\it Graphane} was first predicted by Sofo {\it
et.al.}\cite{sofo} on the basis of electronic structure calculations and has been
recently synthesized by Elias {\it et.al}\cite{elias}. Their experimental work also
showed that the process of hydrogenation is reversible,  making {\it graphane} a potential
candidate for hydrogen storage materials.  Since upon hydrogenation, graphene, a semi-metal
turns into an insulator, it is a good candidate for investigating the nature of metal
insulator transition (MIT).  It may be noted that MIT is a widely studied area in
condensed matter physics \cite{abrahams}.

There are only a few reports on the properties of partially hydrogenated graphene, mainly
focusing on the interaction of a small number of hydrogen atoms with monolayer
graphene.~\cite{boukhvalov,casolo,denis} Electronic structure of hydrogen adsorbed on
graphene has been investigated using density functional theory (DFT) by Boukhvalov {\it et
al}\cite{boukhvalov} and Casolo {\it  et al}.\cite{casolo} Their results support the use
of graphene as the hydrogen storage material. Their results also indicate that the
thermodynamically and kinetically favoured positions of hydrogen are those that minimize
sublattice imbalance. A recent work explores the role of H-frustration in {\it graphane}
like structure using reactive classical molecular dynamics.\cite{flores} On the basis on
DFT Zhou {\it et al} \cite{zhou} proposed a new ordered compound - graphone - obtained by
50 \% hydrogenation showing ferromagnetic behaviour with high Curie temperature.  Other
reports on {\it graphane} include the study of single hydrogen impurity using GW method
\cite{GW}, electronic structure on one and two vacancies in {\it graphane} \cite{pujari}
and the properties of {\it graphane} nanoribbons.\cite{gnr} However the nature of MIT in
graphene-{\it graphane} system remains to be investigated.

In the present work, we investigate the nature of MIT by carrying out a detailed {\it ab
initio} DFT investigation of hydrogenated graphene. The focus of our work is to understand
and gain insight into the way band gap opens.  Therefore we have carried out the
electronic structure calculations for fifteen different hydrogen concentrations between
graphene  and {\it graphane}. Our calculations reveal that, graphene first turns into a
metal with nearly constant density of states (DOS) around the Fermi level then into an
insulator.  Remarkably, the metallic phase is spatially inhomogeneous in the sense, it
consists of insulating islands containing hydrogenated carbon atoms surrounded by the
conducting channels formed by the naked carbon atoms.

\begin{figure*}
  \begin{center}
    \includegraphics[width=4.7cm]{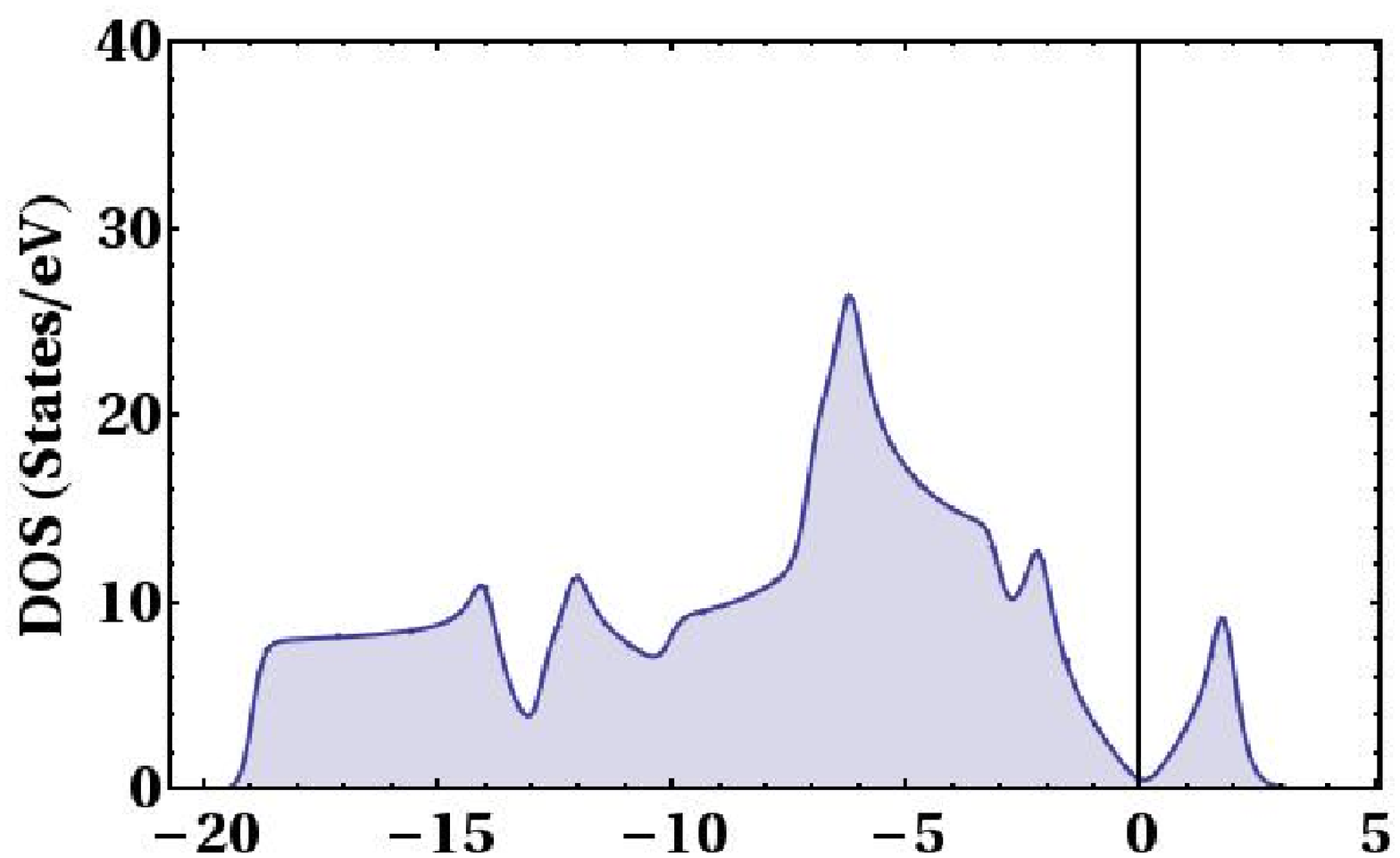}
    \includegraphics[width=4.5cm]{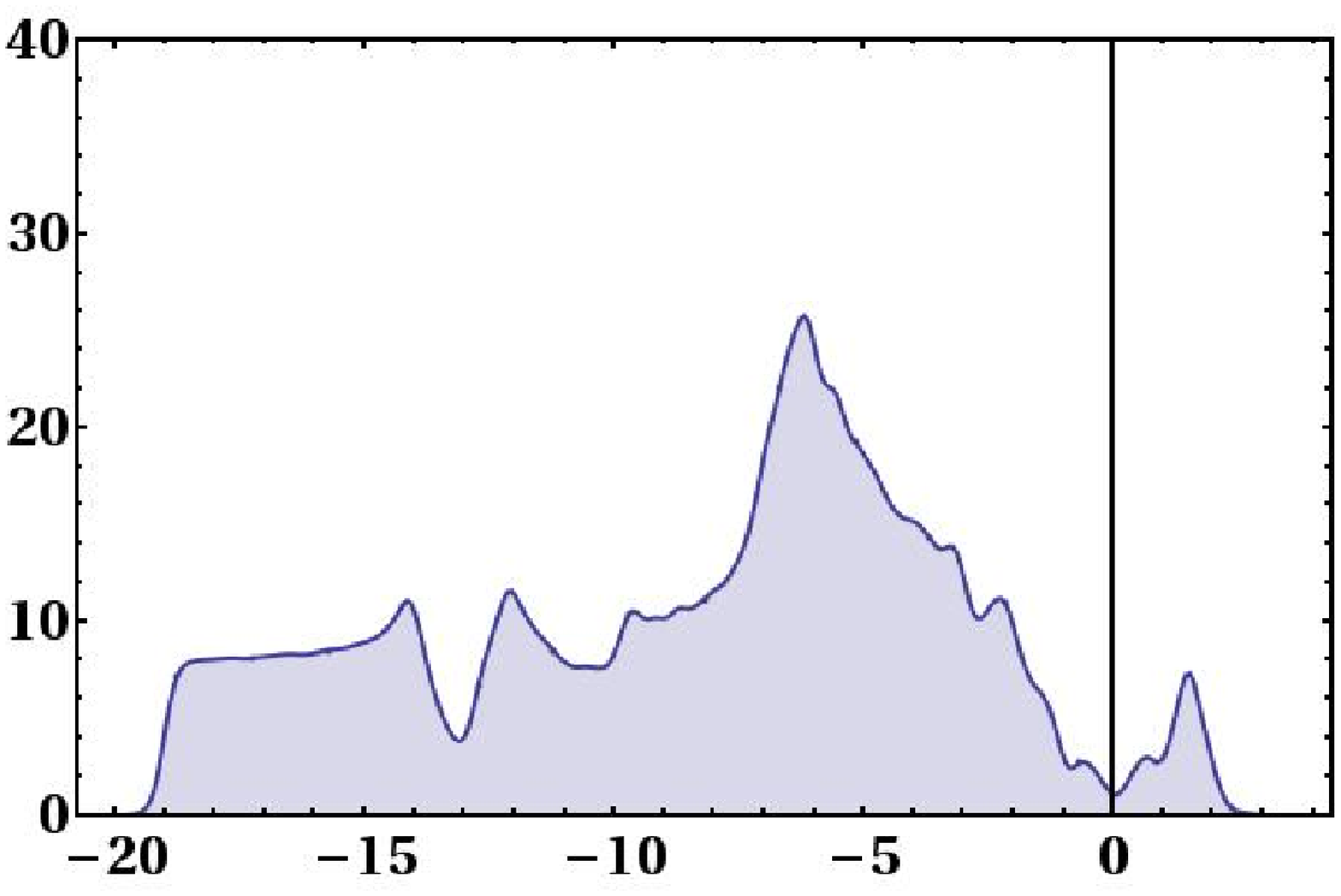}\\
    (a) Graphene \hskip 3cm (b) 8\% H\\~\\
    \includegraphics[width=4.7cm]{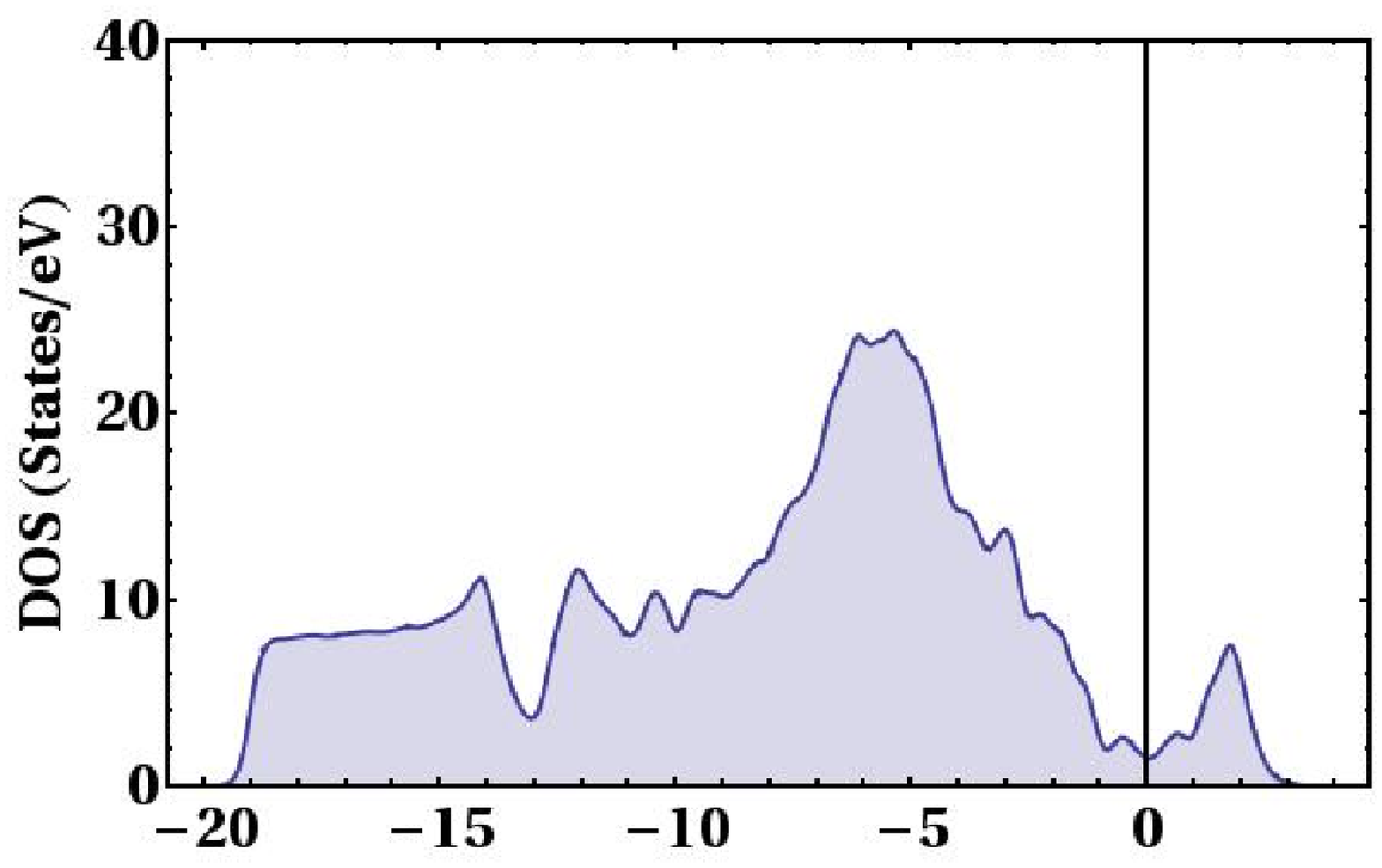}
    \includegraphics[width=4.5cm]{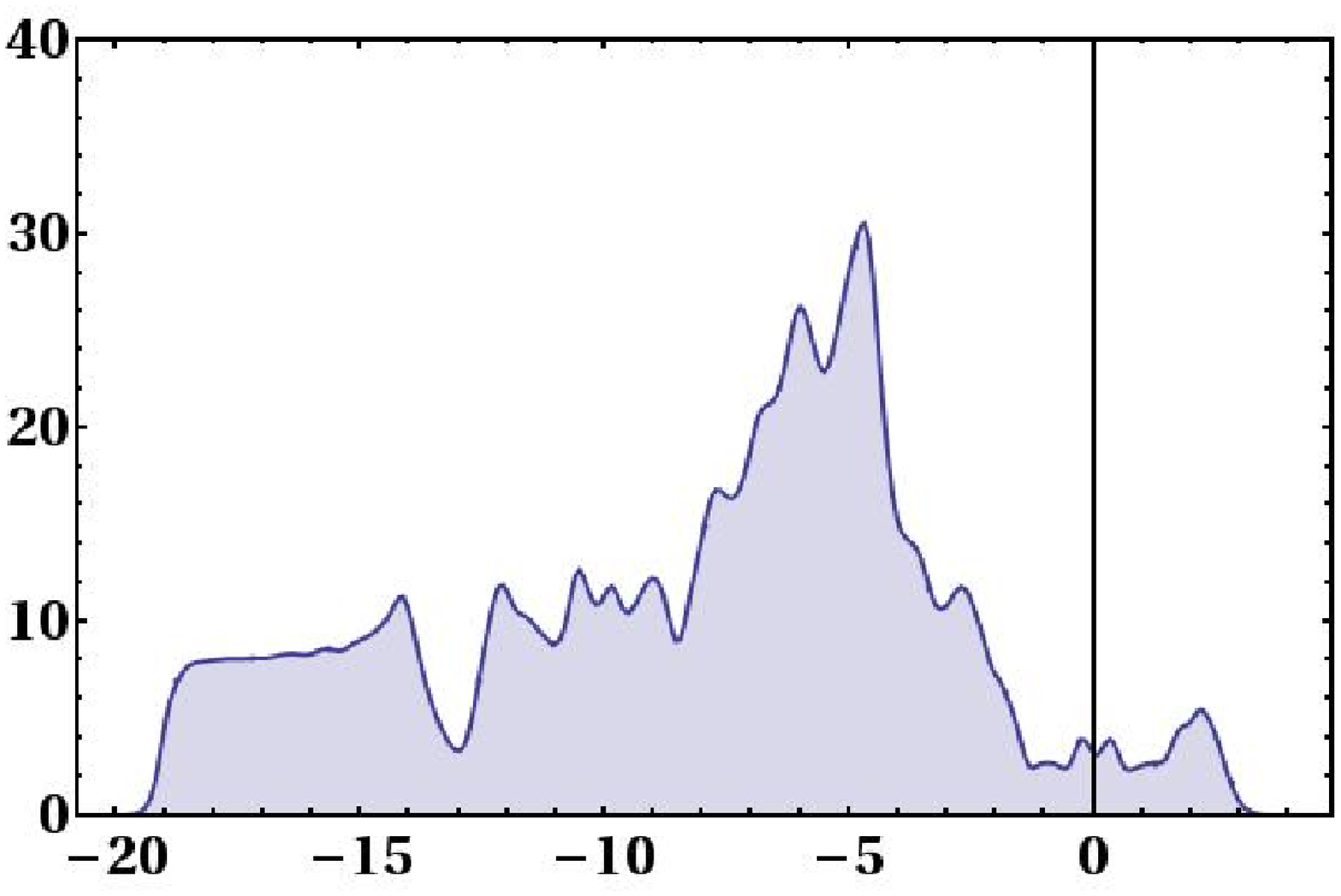}\\
    (c) 20\% H \hskip 3cm (d) 40\% H\\~\\
    \includegraphics[width=4.7cm]{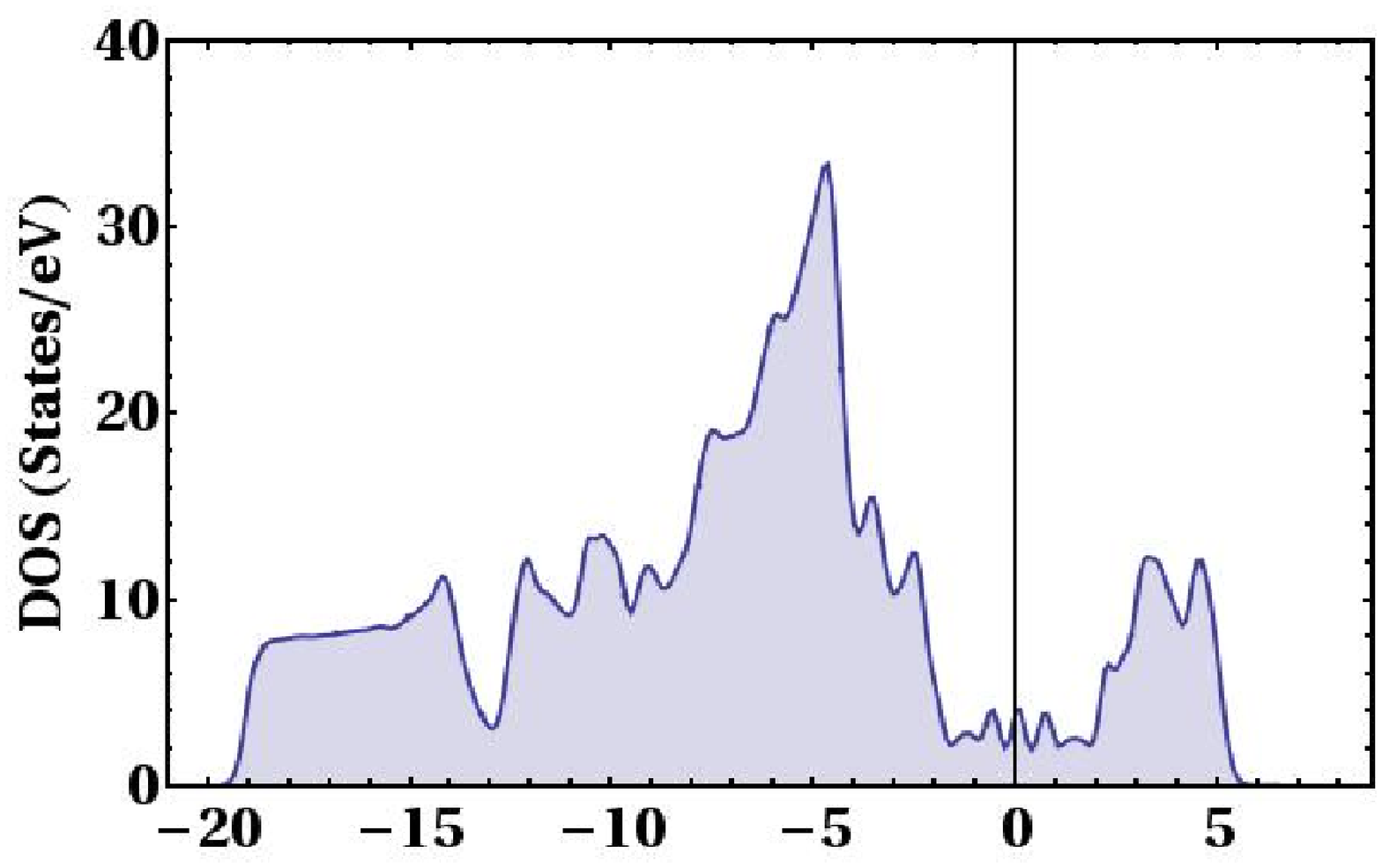}
    \includegraphics[width=4.5cm]{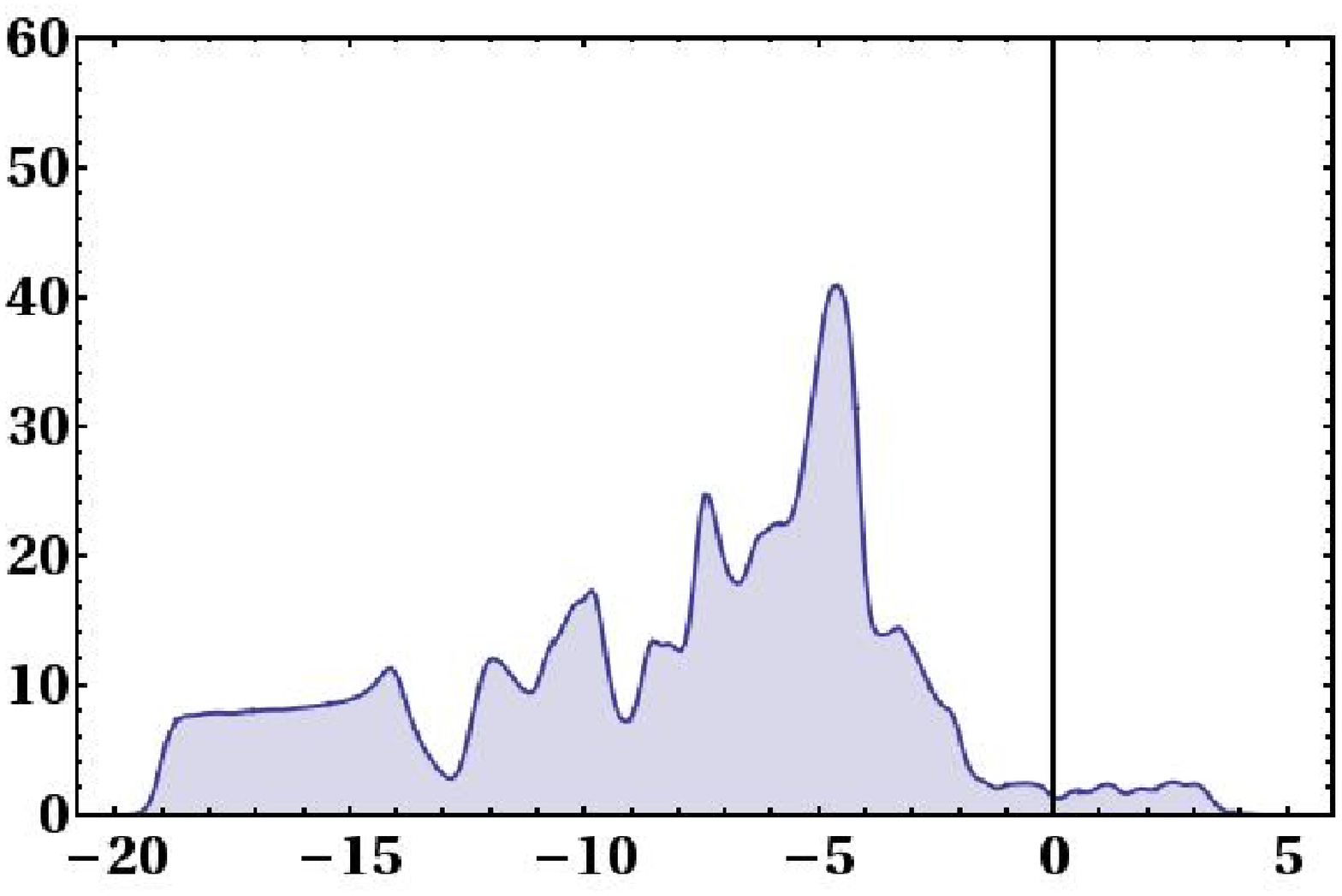}\\
    (e) 50\% H \hskip 3cm (f) 70\% H\\~\\
    \includegraphics[width=4.7cm]{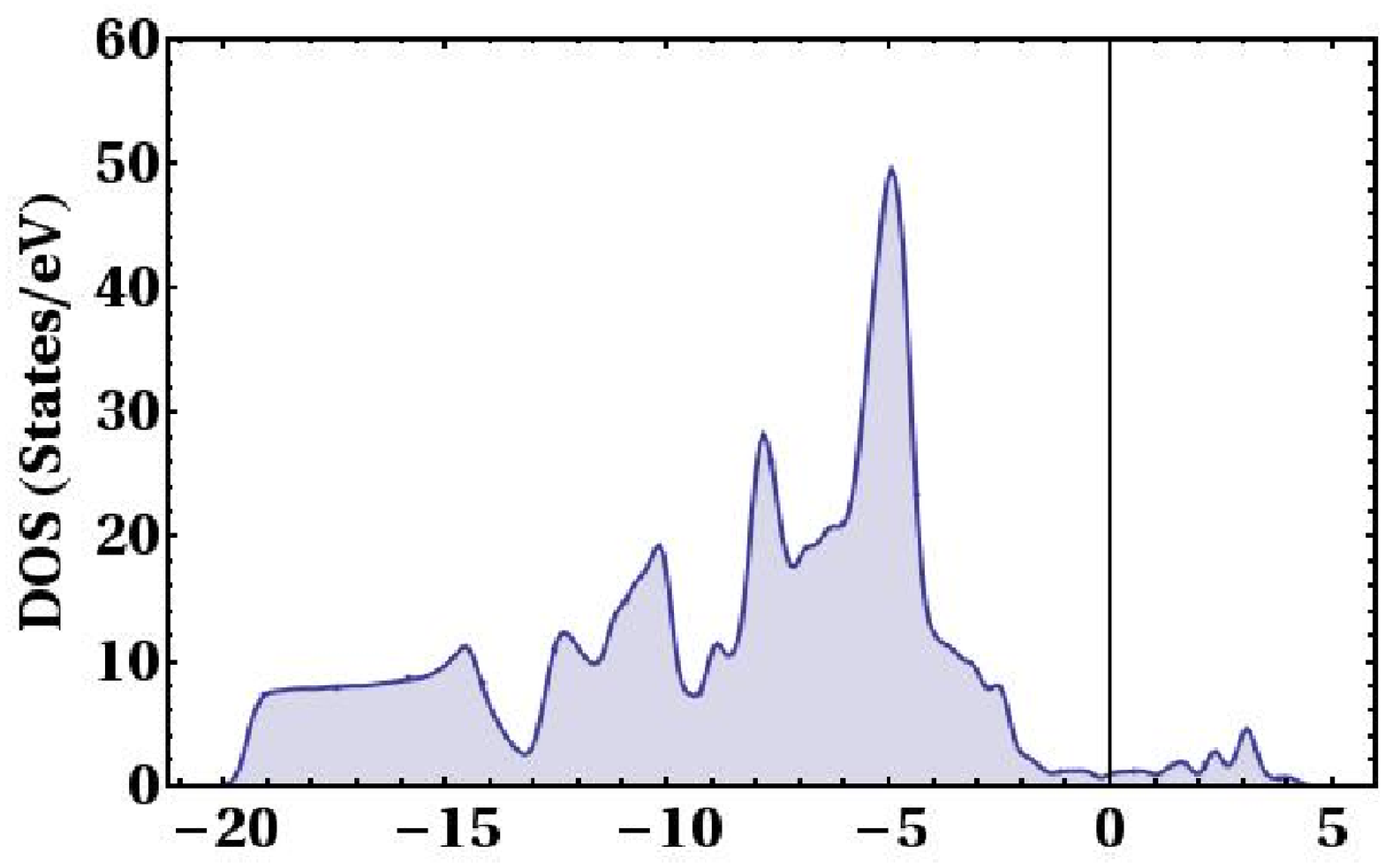}
    \includegraphics[width=4.5cm]{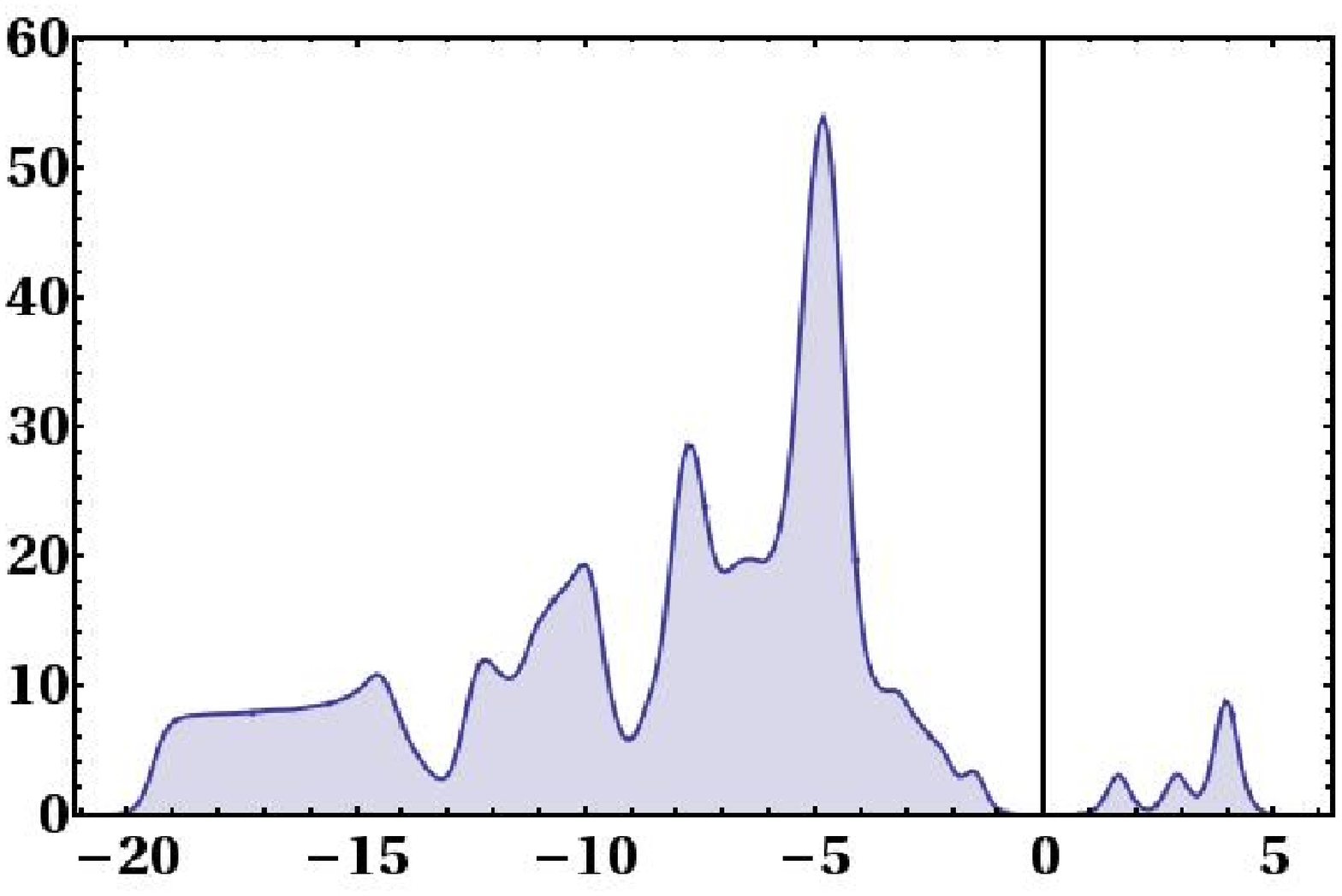}\\
    (g) 80\% H \hskip 3cm (h) 92\% H\\~\\
    \includegraphics[width=4.9cm]{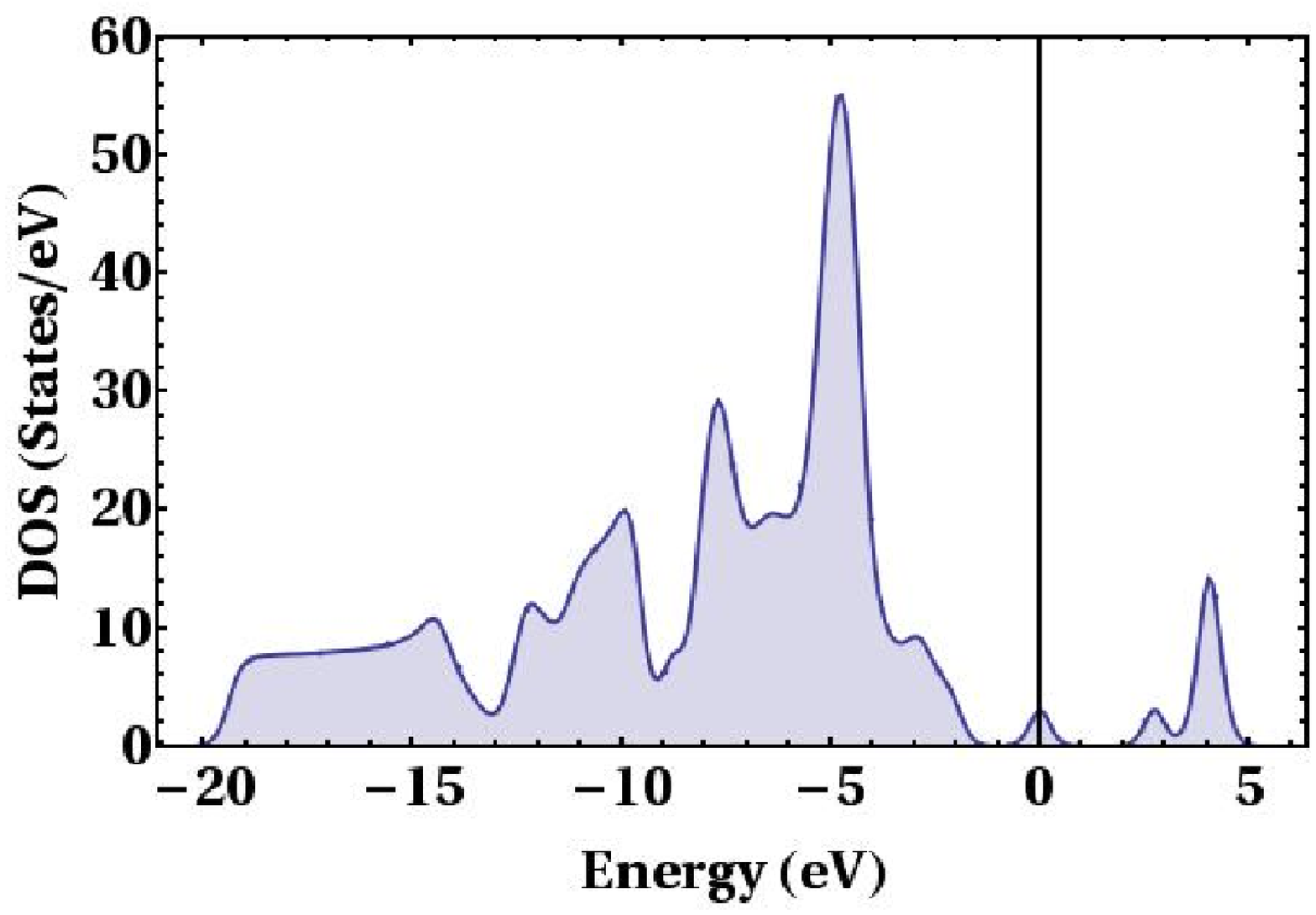}
    \includegraphics[width=4.7cm]{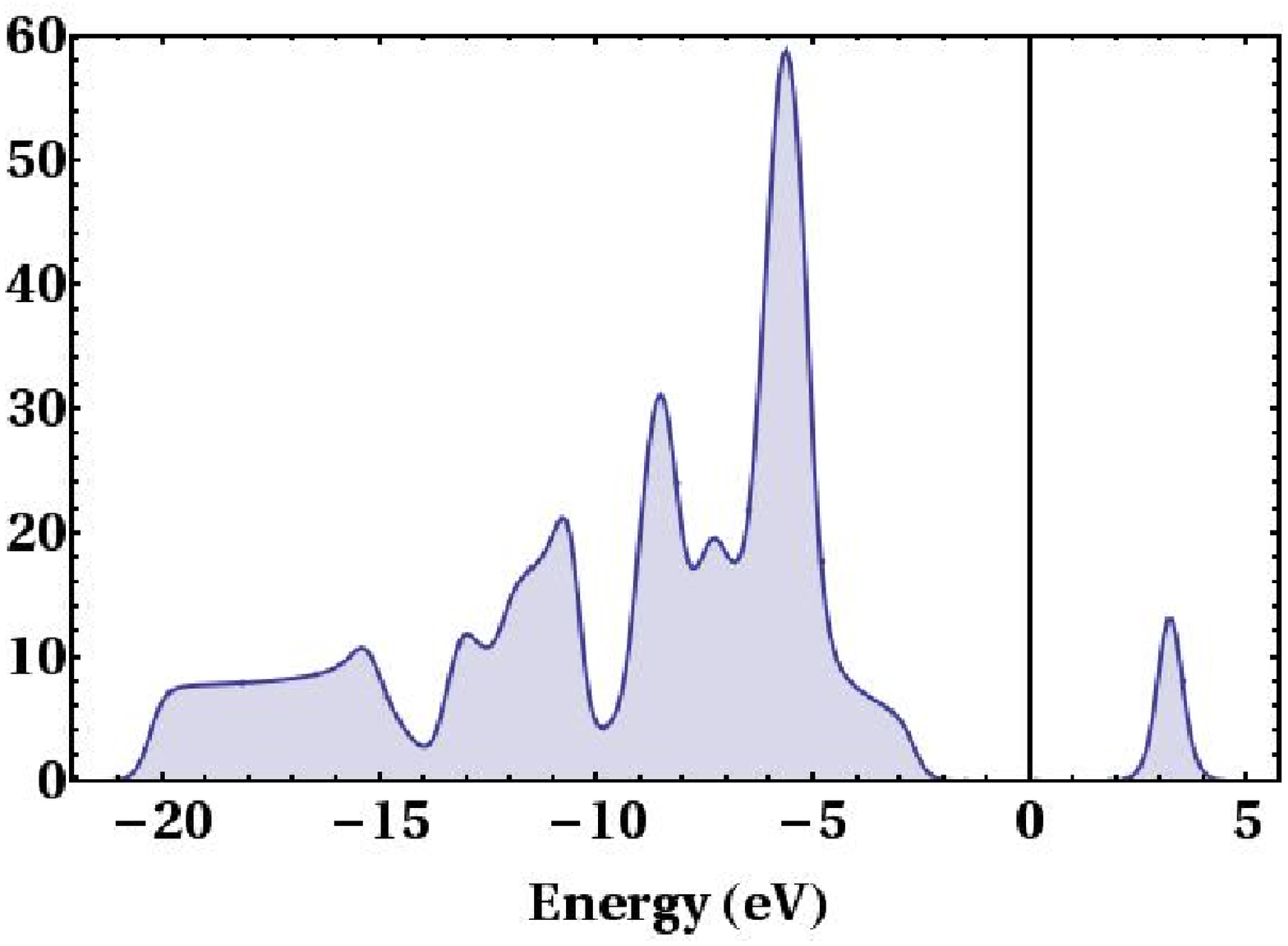}\\
    (i) 96\% H \hskip 3cm (j) {\it Graphane}\\~\\
  \end{center}
\caption{\label{fig1} The total DOS for hydrogenated graphene for various hydrogen
concentrations. The zero of the energy is taken at the Fermi level and is marked by a vertical line}
\end{figure*}

All the calculations have been performed using plane-wave based code VASP~\cite{vasp}
employing projected augmented wave method.  The generalized gradient approximation as
proposed by Perdew, Burke and Ernzerhof \cite{gga1,gga2} has been used for the exchange
correlation potential.  We have used a large unit cell containing 50 carbon atoms for the
hydrogen coverage up to 50\% and  72 carbon atoms for the higher hydrogen coverages.  The
placement of hydrogen is based on chair conformer.\cite{sofo} The convergence criterion
used for the total energy and the forces are 10$^{-5}$ eV and 0.005 eV/{\AA} respectively.
It was found that at least 9 $\times$ 9 {\bf $k$} grid was required during geometry
optimization for acceptable convergence. However, a minimum of 17 $\times$ 17 {\bf $k$}
grid was necessary for obtaining  accurate DOS.  In order to decide the minimum energy
positions for hydrogen atoms, following procedure was adopted. Up to 20\% coverage of
hydrogen, we have carried out the geometry optimization for two different configurations,
first one by placing the hydrogen atoms randomly and second one by placing the hydrogen
atoms contagiously, so as to form a compact island of hydrogenated carbon atoms.  It turns
out that the configuration with the island of the hydrogen atoms is energetically lower by
at least 0.6 eV.  Thus hydrogen prefers to decorate the lattice in contagious and compact
manner. Therefor for the concentrations higher than 20\%, we have decorated the lattice so
as to form the compact island of hydrogenated carbon atoms.


We begin the discussion by presenting the total DOS in Fig. \ref{fig1} for different
hydrogen coverages.  It can be seen that in the low concentration region (less than 20\%),
the essential feature of the DOS for pure graphene namely, the V shaped valley near the
Fermi level is approximately  retained.  As the hydrogen coverage increases there is a
significant increase in the value of DOS at the Fermi level.  The process of hydrogenation
is accompanied by the change in the geometry.  The hydrogenated carbon atoms are moved out
of the graphene plane, the lattice gets distorted and the symmetry is broken. As a
consequence, more and more  ${\mathbf k}$ points contribute to the DOS near Fermi level,
the increase being rather sharp after 20\% coverage.  The region ranging from 30\%
coverage to about 75\% coverage is characterized by the finite value DOS of the order of
2.5/eV or more at the Fermi energy.

\begin{figure}
\centerline{   \includegraphics[width=6cm]{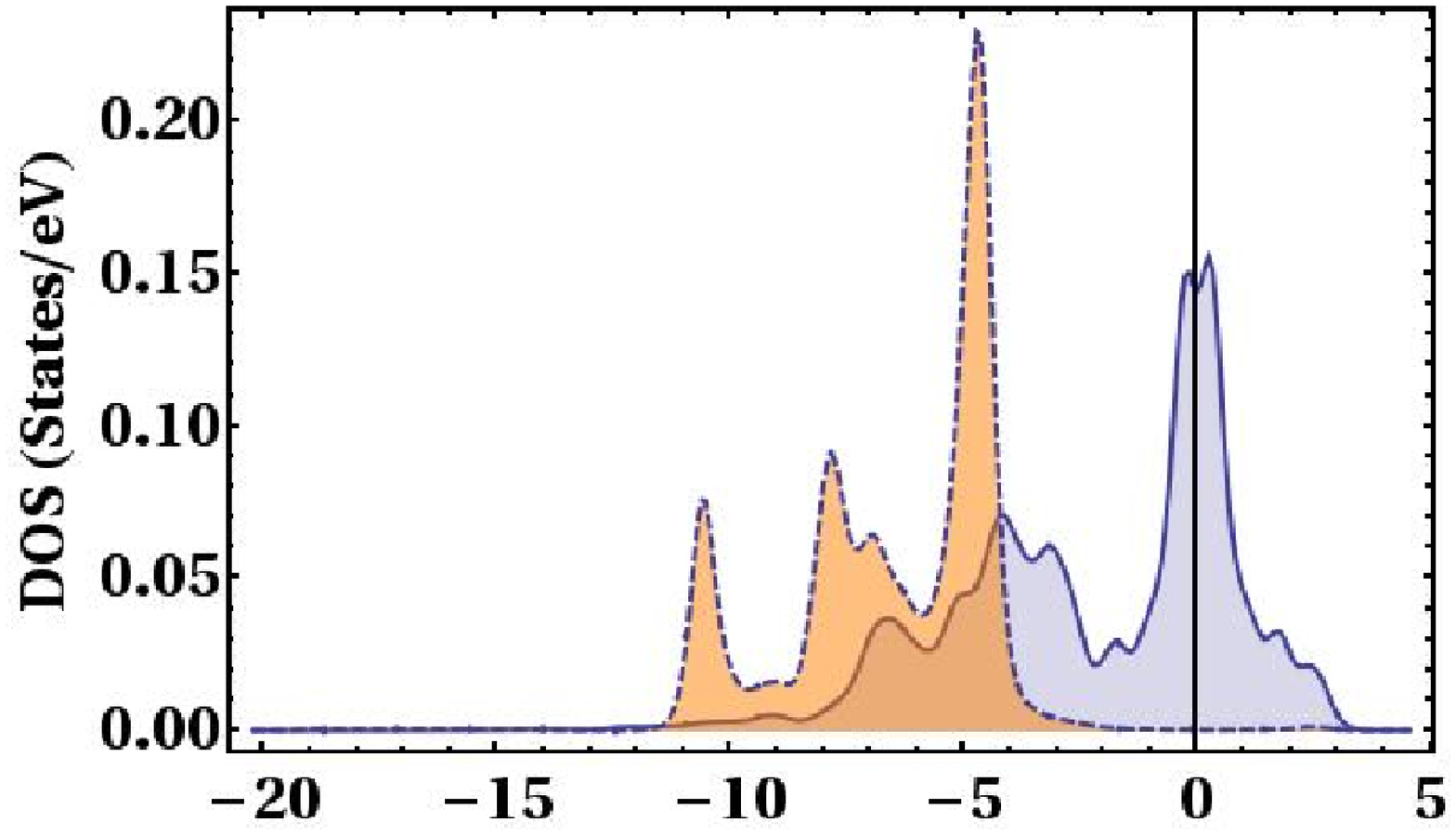}  }
  \caption{Site projected DOS for hydrogen coverage of 40\%. The dotted line indicates 
  hydrogenated carbon and the solid line indicate the naked carbon. Only $p_z$ contribution
  is shown which contributes to more than 90\% of the DOS.}
  \label{pdos}
\end{figure}

We have  analyzed the site projected DOS  for all the cases.  In Fig. \ref{pdos}, we show
site projected DOS for 40\% hydrogen coverage  depicting the contributions from
hydrogenated carbon site and a naked carbon site.  Quite clearly, the only contribution at
the Fermi level comes from the naked carbon sites. It turns out that this is a general feature
for all the systems investigated.  It may be emphasized  that in pure graphene all the
carbon atoms contribute to a single {\bf $k$} point (Dirac point). In contrast to this,
upon hydrogenation although not all the carbon atoms contribute, they do so at many {\bf
$k$} points.

As the concentration increases further (above 80\%), there are too few naked carbon atoms
available for the formation of delocalized $\pi$ bonds.   The value of DOS approache zero
and a gap is established with a few mid gap states.

\begin{figure}
  \begin{center}
    \includegraphics[width=6cm]{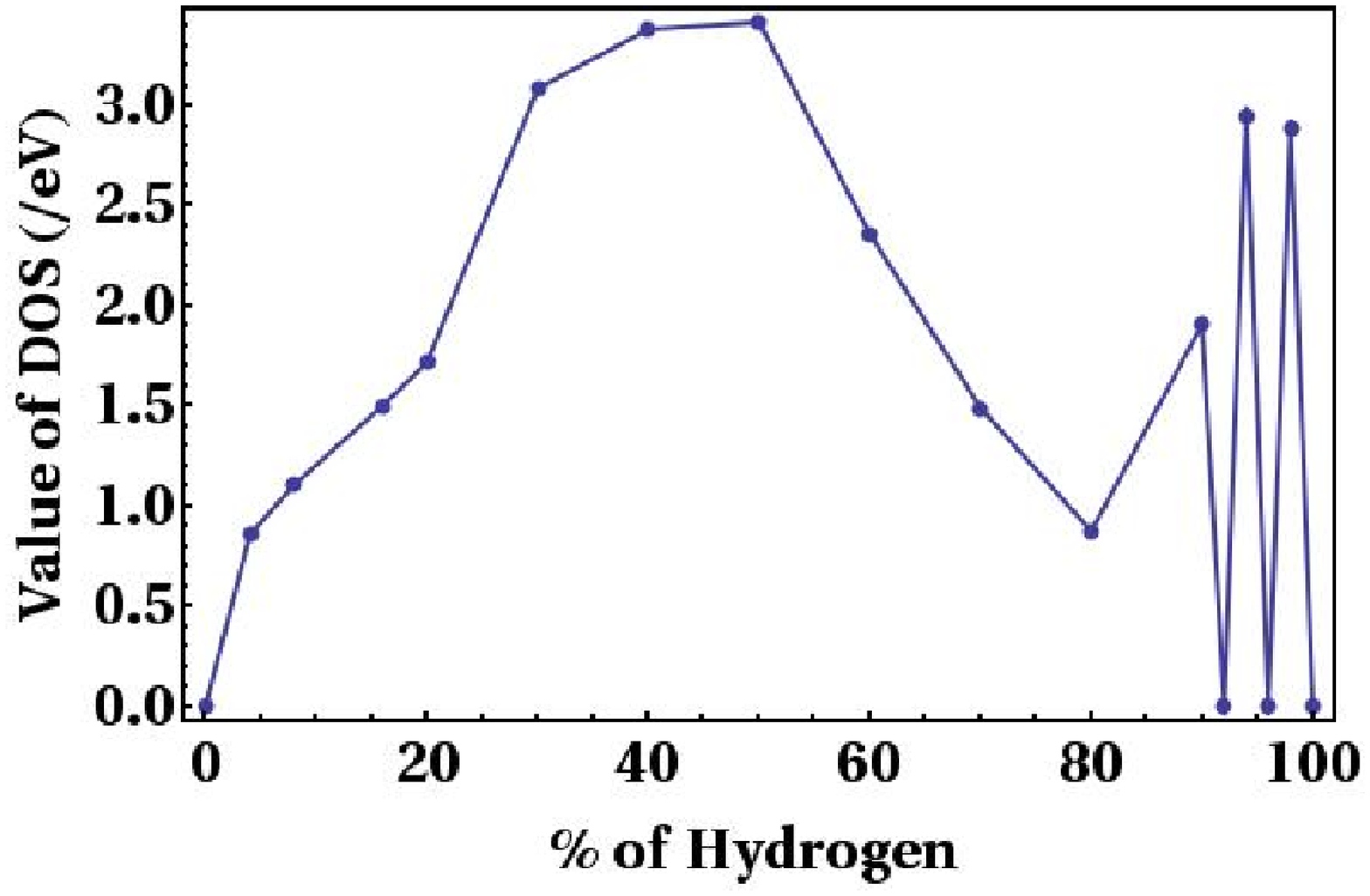}
  \end{center}
\caption{\label{fig3}  Variation of the value of DOS at the Fermi level as a
function of hydrogen coverage.}
\end{figure} 

The evolution towards the metallic state can be better appreciated by examining the
variation in the value of DOS at the Fermi level which is shown in Fig.\ref{fig3}.  A
clear rise in DOS is seen after 20\% hydrogen concentration peaking around 3.5eV at 50\%
concentration. This rise is due to the increasing number of $k$ points contributing to the
Fermi level, as inferred from  analysing the individual bands.  Evidently, over a
significant range of concentration, the value at Fermi level is more than 2/eV. The
decline seen after 60\% is because of the reduction in the number naked carbon atoms.  The
opening of the gap after 90\% is clearly seen.  The system is better described as {\it
graphane} with defects ({\it i.e.} a few hydrogen atoms removed) inducing the mid gap
states. Sudden rise seen in the DOS is precisely due to these states.

\begin{figure}
  \begin{center}
    \includegraphics[width=4.1cm]{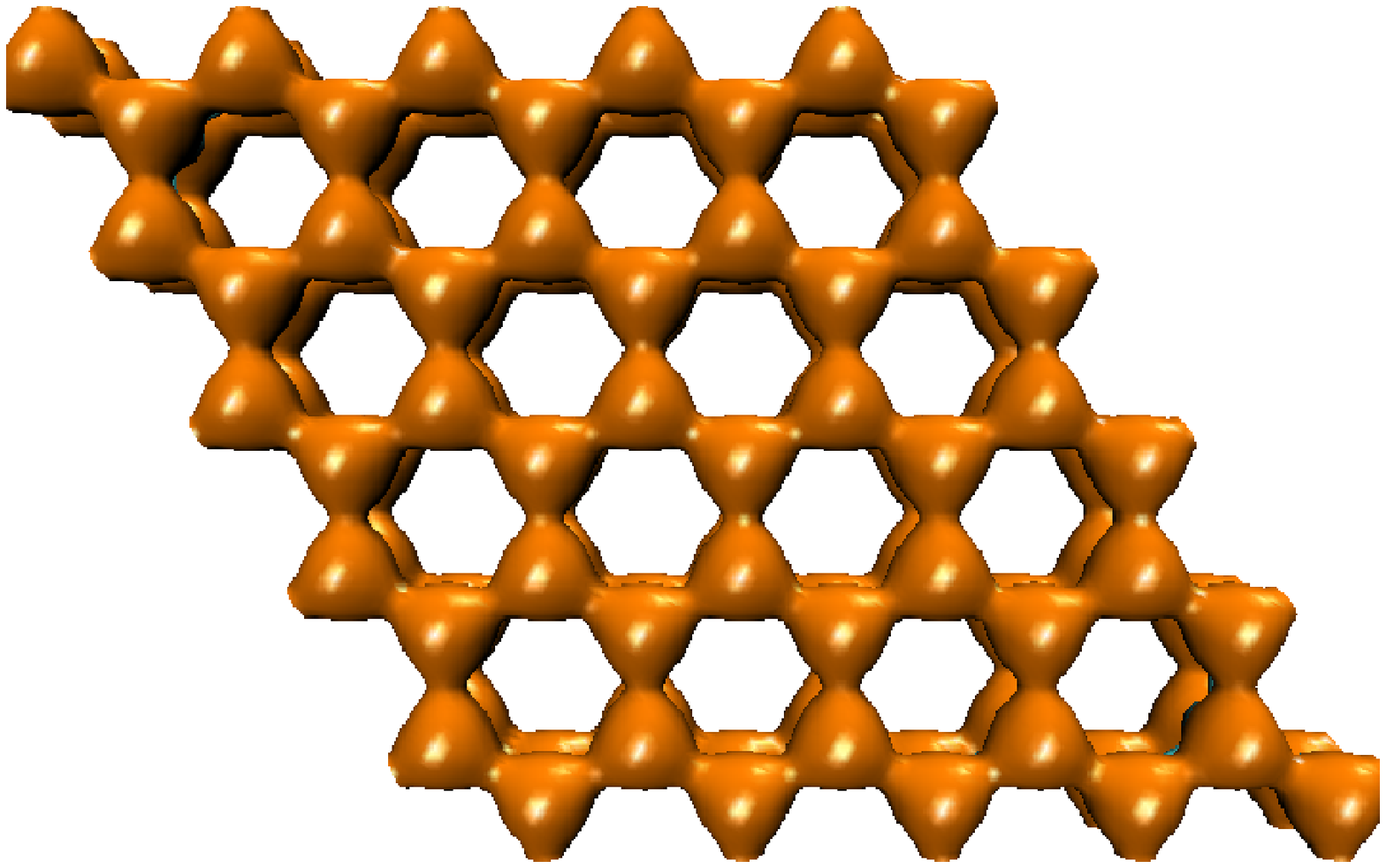}
    \includegraphics[width=4.1cm]{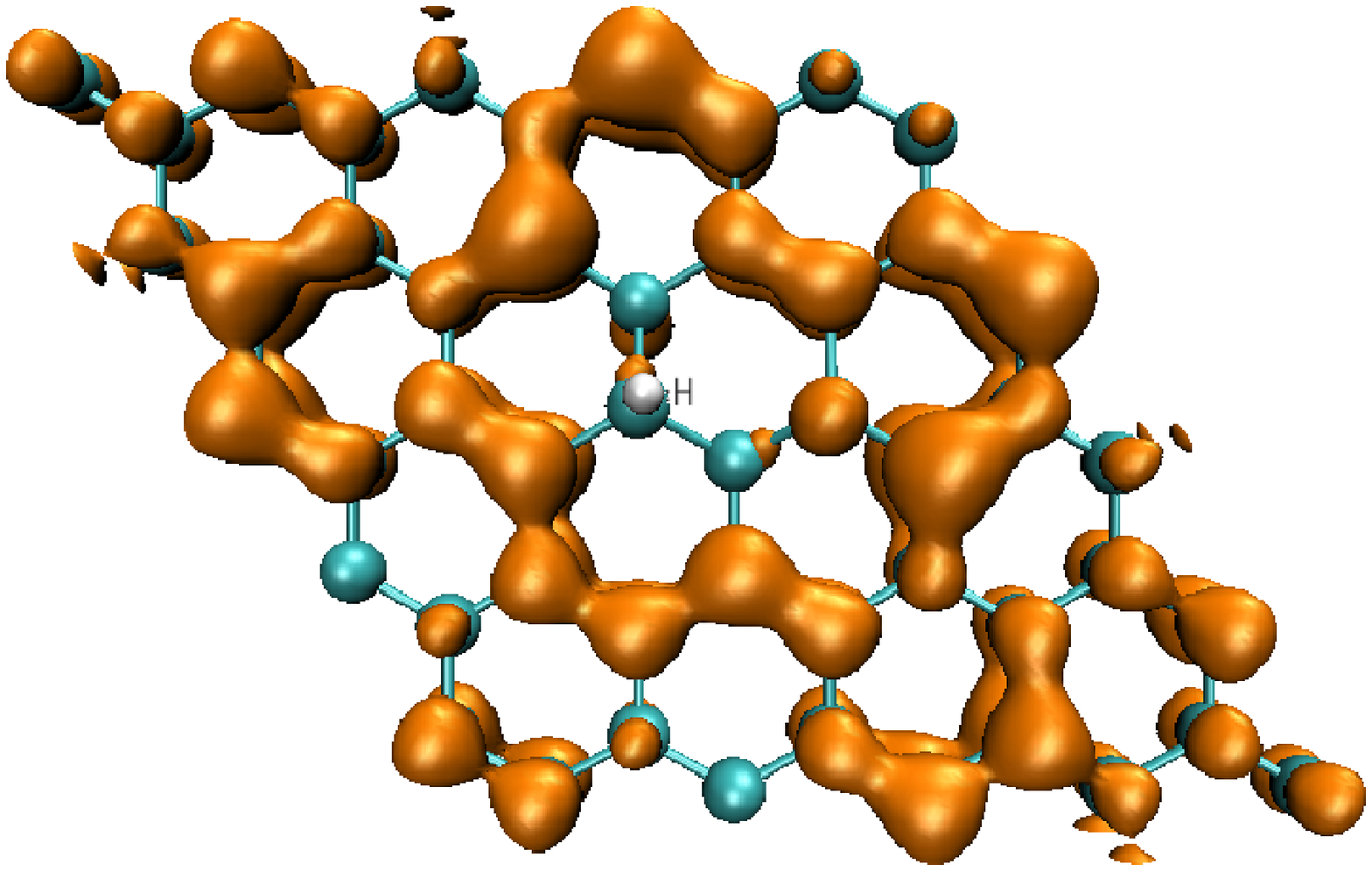} \\
    (a) Graphene \hskip 3cm (b) 8\% H\\~\\
    \includegraphics[width=4.1cm]{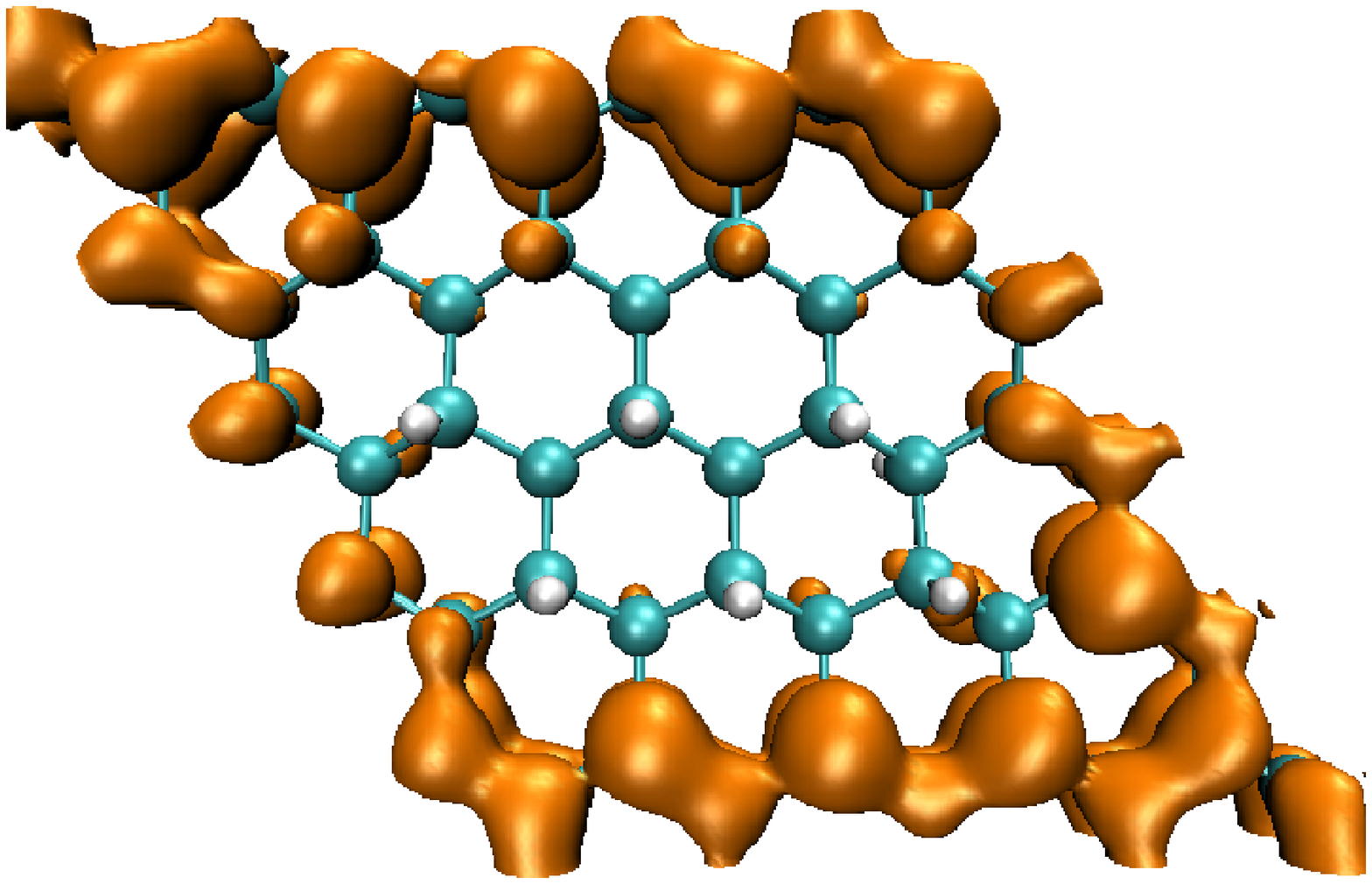}
    \includegraphics[width=4.1cm]{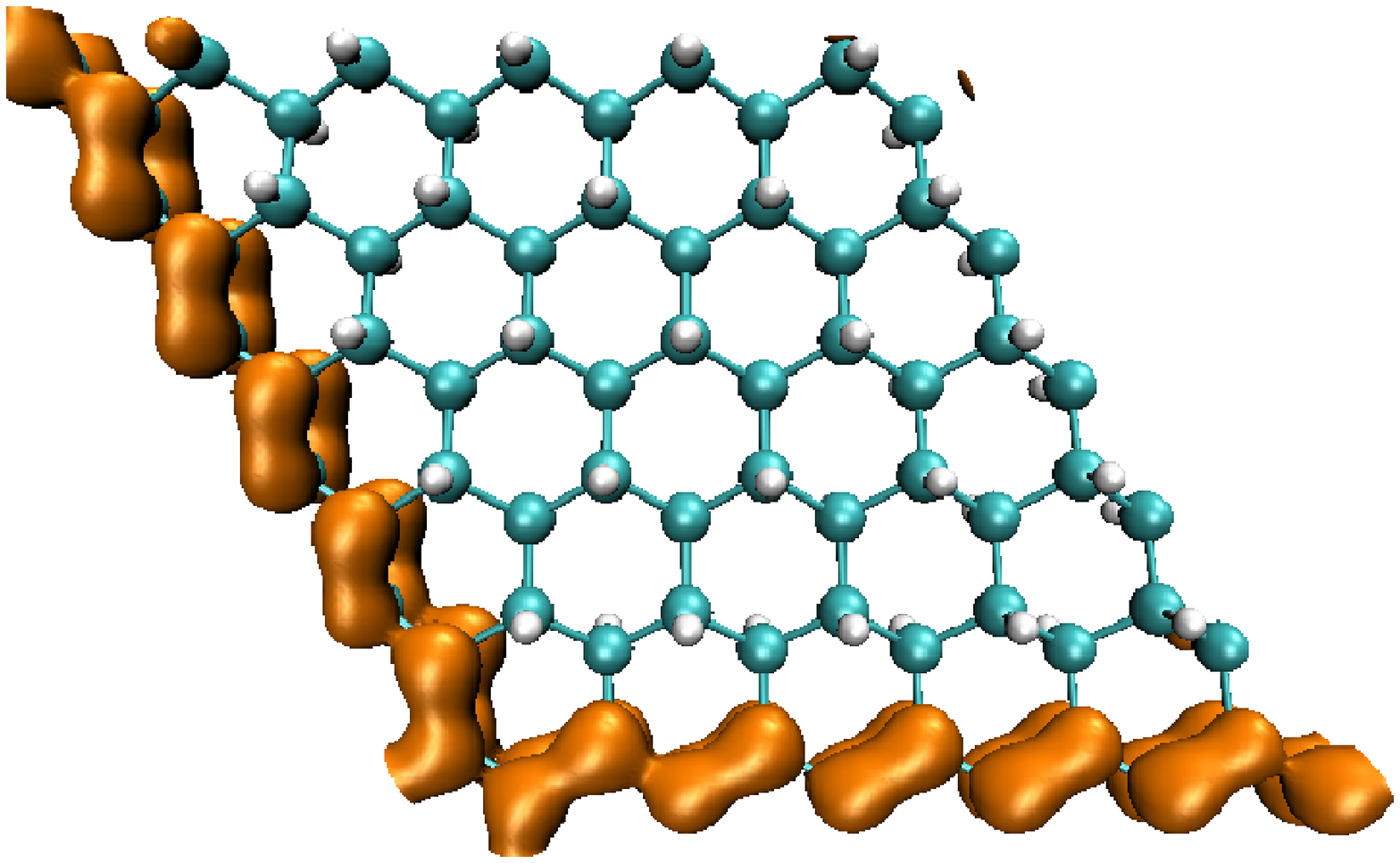}\\
    (c) 40\%H \hskip 3cm (d) 70\% H\\~\\
    \includegraphics[width=4.1cm]{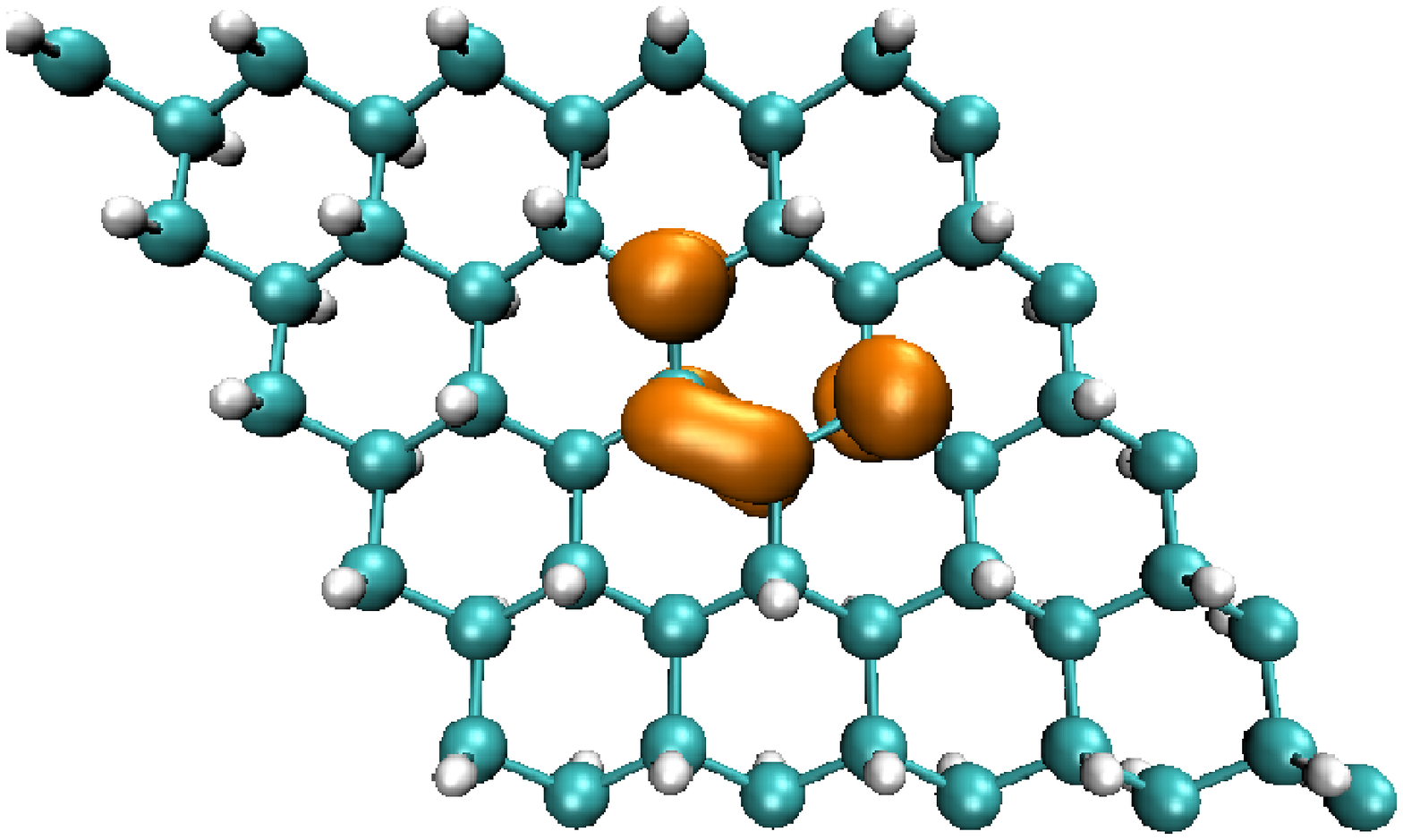}\\
    (e) 92\%H
    \end{center}
\caption{\label{fig4} Isosurfaces of charge densities of bands near the
Fermi level (see text). For comparison the charge density of graphene at Dirac point is
also shown.}
\end{figure} 

Considerable insight can be obtained by examining the evolution of charge densities of the
states contributing near the Fermi level. These are calculated by summing up the
contribution of the bands near the Fermi level from all the  $k$ points.  These are shown
in Fig.(\ref {fig4}) for 8\%, 40\%, 70\% and 92\% hydrogen coverage.  A few interesting
features are immediately evident.  The contrast between the nature of the states for pure
graphene and the states for 8\% coverage is clearly seen.  In the later case, the dominant
contribution comes from delocalized $\pi$ bonded states  formed by $p_z$ orbitals sitting
on naked carbon atoms nearest to the hydrogenated ones.  A particularly striking feature
is the formation of two spatially separated regions as seen in Fig.\ref {fig4}-c and
Fig.\ref{fig4}-d.  The hydrogenated islands do not contribute to the charge density giving
rise to the insulating regions. These are surrounded by $\pi$ bonded naked carbon atoms
forming conducting regions.  This feature is also seen for the higher contributions up to
70\%.  It may be emphasized that the topology in this range of concentrations (30\%-70\%)
shares a common feature namely, there is a contagious region formed by the naked carbon
atoms.  The change in the character of the state at 90\% and above is also evident in
Fig.\ref{fig4}-e.  There are insufficient number of naked carbon atoms to form  contagious
regions.  As a consequence these carbon atoms form localized bonds.  This signifies the
onslaught of the formation of a gap.

In conclusion, our detailed density functional investigations have revealed some
remarkable and novel features of graphene-{\it graphane} MIT.  As
the hydrogen coverage increases, graphene a semi metals, turns first into a metal and then
changes to an insulator. Hydrogenation of graphene pull the carbon atoms out of the plane
breaking the  symmetry of pure graphene. As a consequence many {\bf $k$} points contribute
to DOS at the Fermi level resulting into a metallic case.  The metallic phase has
following unusual characteristic : the sheet shows two distinct regions, a conducting
region formed by naked carbon atoms and embedded into this region are the non conducting
islands formed by the hydrogenated carbon atoms.  The onslaught of the insulating state occurs
when there are insufficient numbers of naked carbon atoms to form connecting channels.
This also means that the transition to insulating phase depends on the distribution of
hydrogen and will occur when the continuous channels are absent.  For a significant range
of hydrogen coverage the DOS is approximately constant near the Fermi level, a
characteristic of nearly free 2D electron gas.  The present work opens up a possibility
of using partially hydrogenated graphene, with designed patterns of the conducting
channels along with the insulating barriers, for the purpose of device applications.

  B.S.P. would like to acknowledge  CSIR, Govt. of India for financial support (No:
  9/137(0458)/2008-EMR-I).  Some of the figures are generated by using VMD software
  \cite{vmd}. P. C. acknowledges M. Arjunwadkar for encouragement and discussion.

 \bibliographystyle{unsrt} 
 \bibliography{biblio}

\end{document}